\documentstyle[twocolumn,seceq]{jpsj}

\newcommand{\diff}[2]{\frac{\partial #1}{\partial #2}}
\newcommand{\sgn}{{\rm sign}}
\newcommand{\vep}{\varepsilon}
\newcommand{\vph}{\varphi}
\newcommand{\pF}{p_{\rm F}}
\newcommand{\veck}{{\mib k}}
\newcommand{\vecp}{{\mib p}}
\newcommand{\vecv}{{\mib v}}
\renewcommand{\i}{{\rm i}}
\renewcommand{\Im}{{\rm Im}}

\def\runtitle{Impurity Effects in Strongly Correlated Metals}
\def\runauthor{Hiroaki {\sc Ikeda} and Kazumasa {\sc Miyake}}

\title
{
Impurity Effects in Strongly Correlated Metals:
Large Pressure Dependence of Residual Resistivity of Heavy Fermions
}

\author
{
Hiroaki {\sc Ikeda} and Kazumasa {\sc Miyake}
}

\inst
{
Department of Material Physics, Faculty of Engineering Science,
Osaka University, Toyonaka 560, Japan
}

\recdate
{
}

\abst
{
It is shown on the basis of the Fermi liquid theory that the s-wave scattering
potential due to nonmagnetic impurity is strongly enhanced by the mass
enhancement factor $1/z$.  As a result the impurity potential with moderate
strength, of the order of the bandwidth of conduction electrons, gives
scattering in the unitarity limit in strongly correlated metals as heavy
fermions.  This effect is embodied by large pressure dependence of residual
resistivities in heavy fermion systems, since the pressure decreases the
degree of correlations, which makes the Kondo temperature increase rapidly.
This accounts for the large pressure dependence of the
residual resistivity observed in heavy fermion metals, such as CeInCu$_{2}$,
CeCu$_{6}$, CeAl$_{3}$.
}

\kword
{
Impurity scattering in Fermi liquid, Ward identity, Periodic Anderson model,
unitarity limit, strongly correlated system,
pressure dependence of residual resistivity
}

\begin{document}
\sloppy
\maketitle

A bunch of careful experiments under pressure for heavy fermion metals have
been carried out recently.  Especially, the systematic variations of
resistivity in the normal state have been investigated in
detail~\cite{flouquet,rf:Kagayama1,rf:Kagayama2}.
It is reported for instance that the temperature $T_{\rm max}$, which
corresponds to the maximum of resistivity and is regarded as being
proportional to the Kondo temperature $T_{\rm K}$, tends to increase rapidly
with increasing pressure~\cite{rf:Kagayama2}.
This fact can be understood, on the basis of a periodic Anderson
model~\cite{rf:Yamada}, in such a way that the increase of hybridization
under pressure causes rapid increase of the renormalization
factor $z$ which is proportional to $T_{\rm K}$.

Futhermore, the residual resistivity in many heavy fermions is reported to
decrease drastically under pressure.~\cite{flouquet,rf:Kagayama1}
This decrease is too large to be explained by only an effect of variation of
the density of states due to that of hybridization.  On the other hand,
it has been believed that the renormalization effect associated with large
mass enhancement cancells out leaving the residual resistivity
unchanged~\cite{fukuyama,varma}.  So, such large pressure dependence may be
attributed to the effect that the impurity potential itself is drastically
renormalized due to the many-body effect.  The purpose of this paper is to
show, on the basis of the Fermi liquid theory, that this is the case.

The results are summarized as follows:
1) It is derived on the basis of the Ward identity argument that the s-wave
scattering potential due to nonmagnetic impurity is strongly
enhanced by the mass enhancement factor $1/z$.  This is one of the
characteristic phenonmena specific to strongly correlated systems.
2) The impurity potential with moderate strength gives scattering in the
unitarity limit in strongly correlated metals as heavy fermions.
3) This can account for the large pressure dependence of the
residual resistivity observed in heavy fermion metals, such as CeCuIn,
CeCu$_{6}$, CeAl$_{3}$, since the degree of correlations is decreased by
applying the pressure making the Kondo temperature increase rapidly.

Hereafter we discuss, on the basis of the Fermi liquid theory, the interplay
between many-body effect and s-wave impurity scattering without the so-called
quantum corrections.  Namely, for the vertex correction, we take into account
all the orders of perturbation in short-ranged repulsive interaction but only
the non-crossing terms in the s-wave impurity scattering.
Such a treatment may be valid for the Fermi liquid which is suffered from
scattering of impurities with dilute enough concentration~\cite{rf:AGD}.
In other words, we assume that these impurities do not break down the
framework of Fermi liquid itself but give a quasiparticle near the Fermi level
a finite life time.  At first, we discuss the system consisting of single
component of fermion, and then the periodic Anderson model later.

To begin with, we discuss how the structure of vertex correction of
particle-hole channel is modified by the presence of impurity scattering.
The coherent part of the Green function, which governs physical properties
at low temperatures, is given as
\begin{eqnarray}
\label{eq:GREEN}
  && G(\vecp,\vep)
      =  z / [ \omega-v(|\vecp|-\pF) \nonumber \\
  && \hspace{15mm} -\i \gamma_{\vecp} \sgn(|\vecp|-\pF)
         -\i \dfrac{\sgn(\vep)}{2\tau} ],
\end{eqnarray}
where $z$ is the renormalization amplitude, and $\gamma_{\vecp}$ and
$1/\tau$ are the damping rate of quasiparticle due to the inelastic
scattering by the repulsive interaction and the elastic one by the impurity
potential.  It should be understood that (\ref{eq:GREEN}) is
obtained after the average over configurations of impurities have been taken.
The expression (\ref{eq:GREEN}) smoothly connects with that of the Fermi
liquid theory in the limit $1/\tau \propto n_{\rm imp} \to 0$, $n_{\rm imp}$
being the impurity concentration.
The damping rate of quasiparticle with low energy is determined predominantly
by $1/\tau$ which is finite even in the low-energy limit.

Then, we consider the asymptotic properties of the vertex function
in the particle-hole channel,
$\Gamma(p_1,p_2;p_1+k,p_2-k) \equiv \Gamma(p_1,p_2;k)$, in the limit
$k=(\veck,\omega) \to 0$, which plays an important role in the Fermi liquid
theory.~\cite{rf:AGD}  As in the pure limit, $\Gamma$ satisfies the following
equation:
\begin{eqnarray}
\label{eq:GAMMA}
  && \Gamma(p_1,p_2;k)=\Gamma^{(1)}(p_1,p_2) \nonumber \\
  && \ \ -\i\int\Gamma^{(1)}(p_1,p)G(p)G(p+k)
            \Gamma(p,p_2;k)\dfrac{d^4p}{(2\pi)^4},
\end{eqnarray}
where $\Gamma^{(1)}$ denotes the irreducible vertex part with respect to
particle-hole lines.  We have set $k=0$ in $\Gamma^{(1)}$, since it has no
singularities at $k=0$.  In the pure limit, the singular contribution arises
from the ``anomalous part'' of particle-hole pair, $G^R G^A$, in the integral
in (\ref{eq:GAMMA}).  Other combinations, $G^R G^R$ and $G^A G^A$, gives
regular contributions.  Thus, if $\tau$ is long enough, an explicit form of
the pair $G(p)G(p+k)$ in the neighborhood
of $|\vecp|=\pF, \vep=0$ can be written as
\begin{equation}
  G(p)G(p+k)=A\delta(\vep)\delta(|\vecp|-\pF)+\vph(p),
\end{equation}
where $\vph(p)$ represents the regular part in which we have set $k=0$ in
$\vph(p)$.

The coefficient $A$ can be determined by integrating $G(p)G(p+k)$
with respect to $\vep$ and $|\vecp|$, and is found to be
\begin{equation}
\label{eq:A}
  A=\dfrac{2\pi\i z^2}{v}
    \dfrac{\omega}{\omega-\vecv\cdot\veck+\i\dfrac{\sgn(\omega)}{\tau}},
\end{equation}
where $\vecv$ is the velocity of quasiparticle with momentum $\vecp$.
Important differences from the Fermi liquid in the pure limit are that
there exists the finite life time $\tau$ due to impurity scattering and
the numerator is proportional to $\omega$.
Namely, the coefficient $A$, (\ref{eq:A}), vanishes for both limits,
$k$-limit and $\omega$-limit, in contrast with the case of pure limit where
$A=(2\pi\i z^2/v)\times \vecv\cdot\veck/(\omega-\vecv\cdot\veck)$ remains 
finite in the $k$-limit while it vanishes in the $\omega$-limit.

Thus, under the existence of impurities, the vertex
$\tilde{\Gamma}\equiv\Gamma^\omega = \Gamma^k$ in the neighborhood
of $|\vecp_{i}|=\pF, \vep_{i}=0$ satisfies the following
integral equation
\begin{eqnarray}
\label{eq:GAMMAOM}
  && \tilde{\Gamma}(p_1,p_2)=\Gamma^{(1)}(p_1,p_2) \nonumber \\
  && \hspace{8mm} -\i\int\Gamma^{(1)}(p_1,p)\vph(p)
                   \tilde{\Gamma}(p,p_2)\dfrac{d^4p}{(2\pi)^4},
\end{eqnarray}
which is the same as that for $\Gamma^\omega$ in the pure limit.
Among the diagrams of $\Gamma^{(1)}$, those representing the pure impurity
scattering may be neglected, because they give small quantum corrections,
of ${\cal O}(\pF\ell) \ll 1$ ($\ell$ being the mean free path due to
impurity scattering) It is to be noted that behavior of the vertex function
for finite $||\vecp|-\pF|$ and $|\vep|$ shows a crossover between that of
the Fermi liquid of pure limit at away from the Fermi surface and
impurity-dominated one on the Fermi surface.
It depends on the strength of impurity scattering where the crossover occurs
in $\vecp$- and $\vep$-space.  In order to discuss the rediual resistivity,
for instance, the renormaliztion for the impurity-dominated region should be
used.  This has been pointed out by Langer long ago~\cite{rf:Langer}.

Now we discuss the vertex correction for the impurity potential due to
the many-body effect.  Hereafter, for the sake of simplicity, we restrict
the impurity potential within that with only the s-wave component.
Consider the vertex corrections for the impurity potential as shown
in Fig.\ 1, where lower order diagrams are drawn. As we discussed above,
diagrams of Fig.\ 1(a),(d),(e) and (f) give negligible contributions of
${\cal O}(\pF\ell)$. Diagrams of Fig.\ 1(g) and (h) give also a small
correction to $\Gamma^{(1)}$ of ${\cal O}(\pF\ell)$.
Thus, the vertex correction arises predominantly from the diagram of the type
Fig.\ 1(b) and (c), and its higher order terms.
It is seen easily in these diagrams that the scattering component with the
higher angular momentum, such as p-, d-wave etc., emerge from the many-body
effect due to the repulsive interaction among electrons,~\cite{rf:Ziegler}
even if the bare impurity potential $u$ has only the to s-wave component.
Nevertheless, the most important scattering process is expected to arise
through the s-wave channel, because the partial-wave phase shifts satisfy the
Friedel sum rule and the phase shift with higher partial wave gives less
contribution in gerneral for local perturbation.  Since the scattering
probability of s-wave does not depend on the momentum transfer, it is
estimated by that of the forward scattering process.

Then, we can estimate the vertex correction for s-wave impurity scattering
with the use of the Ward-Takahashi identity~\cite{rf:Nozieres}
\begin{eqnarray}
\label{eq:WARD}
  && \Sigma(p+k)-\Sigma(p-k) = \int\{G(q+k)-G(q-k)\} \nonumber \\
  && \hspace{1cm} \times\Gamma^{(1)}(q-k,p+k;p-k,q+k)\dfrac{d^4q}{(2\pi)^4},
\end{eqnarray}
and the relation (\ref{eq:GAMMAOM}): The renormalized s-wave scattering
potential $\tilde{u}$ is given by
\begin{equation}
\label{eq:VERTEXC}
  \tilde{u} \equiv \left[1+\tilde{\Gamma}(p,p)\right]u=
   \biggl(1-\left. \diff{\Sigma(p)}{\omega}\right|_{\omega=0}\biggr)u
\end{equation}
This renormalization is represented by the Feynman diagrams as in Fig.\ 2.
The self-energy $\Sigma(p)$ in (\ref{eq:VERTEXC}) includes the effect of
impurity scattering.  However, its effect may be small for the case with small
impurity concentrations where the quantum corrections, of ${\cal O}(\pF\ell) 
\ll 1$, can be neglected.  Thus eq.(\ref{eq:VERTEXC}) is approximated by
\begin{equation}
\label{eq:VERTEXC2}
  \tilde{u}=\frac{1}{z}u,
\end{equation}
where $z$ is the renormalization factor for the Fermi liquid.  The relation
(\ref{eq:VERTEXC2}) has been given by Kotliar {\it et al} without detailed
derivation~\cite{kotliar}, and its derivation has been given very briefly by
the present authors in the theory of anisotropic semiconductiors or semimetals
of heavy fermions~\cite{rf:Ikeda}.  The relation (\ref{eq:VERTEXC2}) can be
interpreted in such a way that the quasiparticles feel the bare potential $u$,
because the potential which the quasiparticle fell is multiplied by $z$, the
weight of quasiparticle in the bare single-electron state.

In strongly correlated system, such as heavy fermions, $z$ is far less than 1
and inversely proportional to the mass enhancement.  Therefore,
the renormalized impurity potential $\tilde{u}$ can become very strong
leading to the scattering in the unitarity limit, even if the bare potential
$u$ is moderate one.

Next, we proceed to calculate the life time $\tau$.  The self-energy
$\Sigma_{\rm imp}(p)$ due to the renormalized s-wave
scattering potential above is given in the $t$-matrix approximation by
\begin{equation}
\label{eq:SIGMA}
  \Sigma_{\rm imp}(\omega)=n_{\rm imp}
    \dfrac{\tilde{u}}{1-\tilde{u}\sum_\veck G(\veck,\omega)}.
\end{equation}
The life time $\tau$ is related with $\Im\Sigma_{\rm imp}(\omega)$ as
$1/2\tau=-z\times \Im\Sigma_{\rm imp}(\omega)$.  So, in order to determine
$\tau$, eqs.(\ref{eq:GREEN}) and (\ref{eq:SIGMA}) need to be solved
self-consistently, in general, as in the case of heavy fermion
superconductors~\cite{rf:Schmitt-Rink,rf:Hirschfeld}
and anisotropic Kondo insulator.~\cite{rf:Ikeda}  However, in the case where
$\tilde{N}(\omega)$, the density of states of quasiparticles in the pure limit,
does not vanish for $\omega=0$, the summation over $\veck$ in (\ref{eq:SIGMA})
is approximately given as
\begin{equation}
\label{eq:DOS}
\sum_\veck G(\veck,\omega)\simeq -i\pi z\tilde{N}(\omega),
\end{equation}
where the particle-hole symmetry is assumed for quasiparticle dispersion.
Thus, the life time $\tau$ is given by
\begin{equation}
\label{eq:TAU}
  \dfrac{1}{\tau} =
    \dfrac{2z\pi\tilde{u}^2 z\tilde{N}(\omega)n_{\rm imp}}
          {1+(\pi\tilde{u}z\tilde{N}(\omega))^2}.
\end{equation}
Here, we should note that in the unitarity limit,
$\tilde{u}z\tilde{N}(\omega)\gg 1$, $1/\tau\simeq 2/\pi\tilde{N}(\omega)$
does not depend on the impurity potential at all.

Now, to discuss the residual resistivity of Ce-based heavy fermions, let us
consider the periodic
Anderson model with a small amount of impurities.  Here the bare impurity
potential for conduction electron is different from that for $f$-electron
in general.  However, we neglect the difference for a moment.
Repeating the above discussions, it is shown that the impurity
potential for the $f$-electrons is enhanced by a factor
$1/z_f\simeq 1-\partial\Sigma_{f}(p)/\partial\omega$, while the one
for conduction electrons is not.  Since the impurity potential for conduction
electron is not renormalized and the quasiparticles near the Fermi level is
composed almost from $f$-electrons~\cite{rf:Yamada},
the effect of impurity scattering comes predominantly from impurity potential
acting on the $f$-electron provided that the bare impurity potential is not
very strong.  This may justify to neglect the difference between impurity
potential for conduction electron and $f$-electron.

Then the life time of the quasiparticle near the Fermi level is
given by the expression similar to (\ref{eq:TAU}):
\begin{equation}
\label{eq:TAU2}
  \dfrac{1}{\tau} =
    \dfrac{2z_f\pi\tilde{u}^2 z_f\left< A_{f} \right>
           \tilde{N}(\omega)n_{\rm imp}}
          {1+(\pi\tilde{u}z_f\left< A_{f} \right>\tilde{N}(\omega))^2},
\end{equation}
where $\left<A_{f}\right>$, the average weight of $f$-electron contained in
the quasiparticle state on the Fermi surface, is given as
\begin{equation}
\label{eq:structure}
A_{f}=\frac{1}{2}\biggl[1-{\tilde{\epsilon}_{f}-\xi_\veck\over
     \sqrt{(\tilde{\epsilon}_{f}-\xi_\veck)^2+4z_f|V_\veck|^2}}
     \biggr],
\end{equation}
where the notations are standard ones~\cite{rf:Ikeda}.
In the strongly correlated limit, $\left<A_{f}\right>\simeq 1$ as mentioned
above.

The residual resistivity $\rho_{0}$ is given by the Drude formula as
\begin{equation}
\label{eq:rho}
\rho_{0}={m^*\over ne^2}{1\over \tau},
\end{equation}
where $m^*$ is the effective mass of hybridization band enhanced by
$1/z_f$.  Substituting (\ref{eq:TAU2}) into (\ref{eq:rho}),
the $z_f$-dependence of $\rho_{0}$ is given as follows:
\begin{equation}
\label{eq:rho2}
\rho_{0}\propto{u^2\left< A_{f} \right>N(\omega)\over
                 z_f^2+(\pi u\left< A_{f} \right>N(\omega))^2},
\end{equation}
where $N(\omega)=z_f\tilde{N}(\omega)$ is the density of states of
hybridyzation band without the repulsive interaction.
It is to be noted that $\rho_{0}$, (\ref{eq:rho2}), does not depend on the
renormalization factor $z_f$ in the strongly correlation
limit~\cite{fukuyama,varma}, $z_f\ll 1$.
However, the $z_f$-dependence in eq.\ (\ref{eq:rho2}) can be
observed if the value of $\tilde{u}z_f\left< A_{f}\right>\tilde{N}(\omega)$
is varied from $\sim 1$ to $\gg 1$.  Indeed, such variations can be realized
by applying hydrostatic pressure, because the mass enhancement
($\propto 1/z_f$)
sensitively depends on the hybridization between $f$-electron and conduction
electron in heavy fermions.

Applying the pressure to feavy fermion metals, such as CeInCu$_{2}$,
CeCu$_{6}$, CeAl$_{3}$, the temperature $T_{\rm max}$
at which the resistivity exhibits maximum increases
rapidly~\cite{rf:Kagayama2}.  This is
considered to be due to the rapid increase of $z_f$.  Correlating with this,
$\rho_{0}$ shows the rapid decrease~\cite{flouquet,rf:Kagayama1,rf:Kagayama2}
which is too rapid to be explained by
the change of band structure under pressure as observed in ordinary metals.
For example, $\rho_{0}$ of CeInCu$_{2}$ decresses from 85 $\mu\Omega$cm at
ambient pressure to 20 $\mu\Omega$cm at $P$=8 GPa~\cite{rf:Kagayama1},
while its $T_{\rm max}$ increases from 27 K to 1000 K
correspondingly~\cite{rf:Kagayama2}.  This behavior is consistent
with the one expected from eq.(\ref{eq:rho2}).
Thus, the large pressure dependence of $\rho_{0}$ observed in heavy fermions
reveals explicitly the way of mass enhancement in which the large frequency
dependence of the self-energy plays a crucial
role~\cite{rf:Yamada,fukuyama,varma,yoshimori,jichu,miyake}.
However, as $z \to 1$, the specific properties of compound, such as anisotropy
of the density of states at the Fermi level and impurity potential with higher
partial wave component beyond s-wave, should be taken into account.

In summary, we have discussed the effect of impurity scattering in strongly
correlated metals on the basis of the Fermi liquid theory.
We have found that the differnce between $k$-limit and $\omega$-limit of the
vertex in the harticle-hole channel disappears due to the damping by the
impurity scattering.  In this case the s-wave scattering process is
enhanced by a factor $1/z$ so that this renormalization process is very
important in the strongly correlated systems where $z\ll 1$.
And then, we have taken into account the effect of this renormalized
impurity scattering through the $t$-matrix approximation.
As a result, the residual resistivity $\rho_{0}$ shows a large dependence on
$z$ as $z \to 1$.  This is consistent with the large pressure dependence of
$\rho_{0}$ observed in many heavy fermions.

We have much benefitted from informative conversations and correspondences
with G. Oomi and T. Kagayama.
We would like to acknowledge K. Yamada for a critical comment on the
preliminary version of derivation of the vertex correction of impurity
potential, which forced us to clarify its derivation substantially. One of us
(K. M.) would like to acknowledge C. M. Varma for stimulating discussions.
This work is supported by the Grant-in-Aid for Scientific Research (07640477),
and Monbusho International Scientific Program (08044084), and the Grant-in-Aid
for Scientific Research on Priority Areas ``Physics of Strongly Correlated
Conductors'' (06244104) of Ministry of Education, Science and Culture.

\newpage

\begin{description}
\item[Fig.~1 \ \ ]
Diagrams for vertex corrections. The broken line represents the
impurity potential $u$ of $s$-wave component and the wavy line the Coulomb
interaction. The solid line represents the bare Green function. $\Gamma$ is
the full vertex due to the Coulomb repulsion and impurity potential.
\item[Fig.~2 \ \ ]
Vertex correction of impurity potential of s-wave channel.
\end{description}

\begin{thebibliography}{99}
\bibitem{flouquet}
J. Flouquet, P. Haen, P. Lejay, P. Morin, D. Jaccard, J. Schweizer,
C. Vettier, R. A. Fisher and N. E. Phillips:
J. Magn. Magn. Mater. {\bf 90 \& 91} (1990) 377.
\bibitem{rf:Kagayama1}
T. Kagayama, G. Oomi, H. Takahashi, N. Mori, Y. Onuki and T. Komatsubara:
Phys. Rev. B {\bf 44} (1991) 7690;
T. Kagayama, G. Oomi, R. Yagi, Y. Iye, Y. Onuki and T. Komatsubara:
J. Alloys and Compounds. {\bf 207/208} (1994) 271.
\bibitem{rf:Kagayama2}
T. Kagayama and G. Oomi:
J. Appl. Phys. Suppl. {\bf 32} (1993) 318;
T. Kagayama, G. Oomi, E. Ito, Y. Onuki and T. Komatsubara:
J. Phys. Soc. Jpn. {\bf 63} (1994) 3927.
\bibitem{rf:Yamada}
K. Yamada and K. Yosida: Prog. Theor. Phys. {\bf 76} (1986) 621;
K. Yamada, K. Yosida and K. Hanzawa:
Prog. Theor. Phys. Suppl. {\bf 108} (1992) 141.
\bibitem{fukuyama}
H. Fukuyama: {\it Theory of Heavy Fermions and Valence Fluctuations}
eds. T. Kasuya and T. Saso (Springer, Berlin, 1985), p. 209.
\bibitem{varma}
C. M. Varma: {\it Theory of Heavy Fermions and Valence Fluctuations}
eds. T. Kasuya and T. Saso (Springer, Berlin, 1985), p. 277.
\bibitem{rf:AGD}
A. A. Abrikosov, L. P. Gor'kov and I. Ye. Dzyaloshinskiiet:
{\it Quantum Field Theoretical Methods in Statistical Physics}
(Pergamon Press, Oxford, 1965), chap. 4.
\bibitem{rf:Langer}
J. S. Langer:
Phys. Rev. {\bf 124} (1961) 997; Phys. Rev. {\bf 124} (1961) 1003.
\bibitem{rf:Nozieres}
P. Nozi$\acute{\rm e}$res:
{\it Theory of Interacting Fermi Systems}
(Benjamin, New York, 1964), chap. 6.
\bibitem{rf:Ziegler}
W. Ziegler, D. Poilblanc, R. Preuss, W. Hanke and D. J. Scalapino:
\bibitem{kotliar}
G. Kotliar, E. Abrahams, A. E. Ruckenstein, C. M. Varma, P. B. Littlewood
and S. Schmitt-Rink: Europhys. Lett. {\bf 15} (1991) 655.
\bibitem{rf:Ikeda}
H. Ikeda and K. Miyake:
J. Phys. Soc. Jpn. {\bf 65} (1996) No.6 in press.
\bibitem{rf:Schmitt-Rink}
S. Schmitt-Rink, K. Miyake and C. M. Varma:
Phys. Rev. Lett. {\bf 57} (1986) 2575.
\bibitem{rf:Hirschfeld}
P. Hirschfeld, D. Vollhardt and P. W{\"o}lfle:
Solid State Commun. {\bf 59} (1986) 111.
\bibitem{yoshimori}
A. Yoshimori and H. Kasai: J. Magn. Magn. Mater. {\bf 31-34} (1983) 475.
\bibitem{jichu}
H. Jichu, T. Matsuura and Y. Kuroda: Prog. Theor. Phys. {\bf 72} (1984) 366.
\bibitem{miyake}
K. Miyake, T. Mastuura and C. M. Varma: Soild State Commun. {\bf 71} (1989)
1149.
\end{thebibliography}
\end{document}